\documentclass[prl,twocolumn,twoside,showpacs,superscriptaddress]{revtex4}
\usepackage{graphics,amssymb,amsmath,epsf}
\epsfclipon

\begin{document}

\noindent
{\bf Comment on ``Phase Transitions in Systems of Self-Propelled Agents
and Related Network Models''}\\

In a recent letter \cite{ADHKL-PRL},
Aldana {\it et al.} study order-disorder phase transitions 
in random network models and show 
that the nature of these transitions may change with the way noise is 
implemented in the dynamics. 
Arguing that these networks are limiting cases of simple 
models of interacting self-propelled agents of the type of the
Vicsek model (VM) \cite{VICSEK}, they claim 
that the conclusions reached for the networks may carry over to the
transitions to collective motion of the VM-like systems. 
They suggest in particular that in the case of ``angular'' noise
(i.e. as in the original VM \cite{NOTE}, or in their Eq.~(1)) 
the transition to collective motion is {\it continuous}, 
in contradiction with some of the conclusions of \cite{GC-PRL}.
While we agree with the analysis of the network models,
we argue here that it has no bearing on VM-like systems. We show in particular
that the transition to collective motion, for angular noise, 
remains {\it discontinuous} for any finite microscopic velocity $v$ 
and finite density $\rho$, however large, confirming \cite{GC-PRL}.

\begin{figure}
\centerline{
\epsfxsize=8.cm
\epsffile{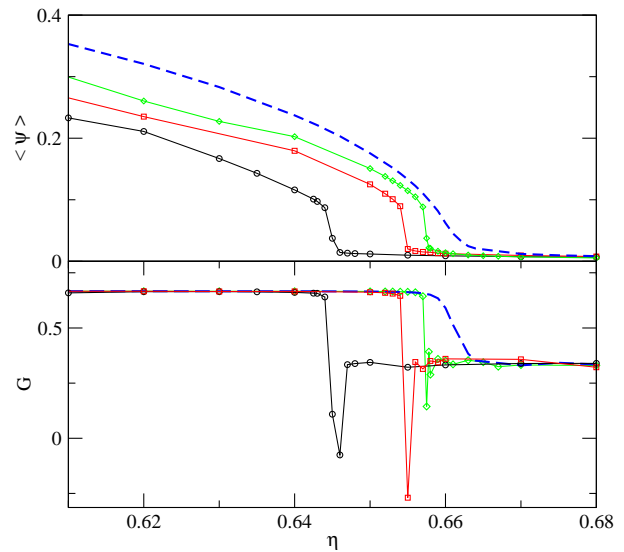}
}
\caption{(Color online)
Order-disorder transition  for the Vicsek model at different $v$
values (solid lines)
and the network of fixed vectors of \cite{ADHKL-PRL} (dashed line),
both with angular noise.
(a) Time-averaged order parameter 
$\langle\psi\rangle$ vs noise strength $\eta$.
(b) same as (a) for the Binder cumulant 
$G=1-\langle\psi^2\rangle^2/\langle\psi^4\rangle$.
For the VM, $\rho=2$, and $v=5$ (circles), $v=10$ (squares), $v=20$ (triangles)
with $L=200$, 250 and 300 respectively.
For the network, the number of nodes is $N=4\times 10^5$ and 
the average connectivity is set to $K=2\pi$ 
(i.e. each node interacts with 6 or 7 neighbors, with suitable probabilities),
which corresponds to $\rho=2$ in the VM.
}
\label{fig}
\end{figure}

In \cite{GC-PRL}, is was shown that the transition in the original VM
appears continuous only when the linear system size $L$ 
is smaller than some crossover size $L^*(\rho,v)$ and thus is 
discontinuous in the thermodynamic limit (see Fig.~2a there).
These results were obtained for values of $\rho$ and $v$ of order unity.
The same scenario occurs in the large-$v$ or large $\rho$ limit 
of interest here.
Figure~\ref{fig}a here displays results akin to Fig.~1 of \cite{ADHKL-PRL}
except that the VM order parameter curves were obtained in systems such 
that $L>L^*(\rho,v)$, whereas in \cite{ADHKL-PRL}  $L\simeq100<L^*(\rho,v)$ 
(see below for our estimates of $L^*$). Clearly, the transition
is discontinuous, as demonstrated by the minimum 
shown by the Binder cumulant  (Fig.~\ref{fig}b) and 
by the bimodality of the order parameter distribution in the transition region
(not shown).

The key difference between the network models and interacting self-propelled 
agents is indeed that the latter move, inducing a local coupling between
order and density, which is well-known to be crucial for understanding
collective properties of active particles \cite{TTR-REV}. 
In the VM, this coupling
gives rise to strong density and order variations on lengthscales of the order
of $L^*$. While the network models in \cite{ADHKL-PRL} 
capture the long-range interactions
due to large velocities, they obviously cannot account for any coupling
between density and order. 
The network models only represent VM-like models of size $L\lesssim L^*$.

The crossover scale $L^*$ is difficult to estimate with high accuracy, 
but our data indicate that $L^*$ increases roughly linearly with $v$
(for $\rho=2$, 
$L^*=150\pm25, 175\pm25, 250\pm50, 550\pm50$ for $v=5,10,20,40$ respectively).
Thus, we expect $L^*$ to be finite at any finite $v$. As a consequence,
the transition is always discontinuous in the thermodynamic limit, although its
asymptotic behavior is harder to observe as $v$ is taken larger and larger.
When $v$ is taken to infinity first, the transition is continuous 
for all finite ``size'', but then the notion of distance in physical space is abolished.
 
Summary: the transition to collective motion in 
VM-like systems with angular noise remains discontinuous for large $v$ values.
Thus, the networks studied in \cite{ADHKL-PRL} at best constitute
a {\it singular} $v\to\infty$ limit of these systems \cite{NOTE2}.

\bigskip
{\obeylines
\noindent Hugues Chat\'e and Francesco Ginelli
{\small CEA --- Service de Physique de l'Etat Condens\'e
91191 Gif-sur-Yvette, France}
\bigskip
\noindent Guillaume Gr\'egoire
{\small Mati\`ere et Syst\`emes Complexes
Universit\'e Denis Diderot, Paris, France}
}


\begin{thebibliography} {10}
\bibitem{NOTE2} In the large $\rho$ limit, we obtain similar results: 
for $L>L^*$ the transition reveals its 
discontinuous character and $L^*$ increases with 
$\rho$ (results obtained at $v=0.5$, not shown).
\end{thebibliography}
\end{document}